\documentclass[epj]{svjour}

\usepackage{graphics}
\usepackage{graphicx}
\usepackage{dcolumn}
\usepackage{bm}

\begin{document}

\title{Socio-economical dynamics as a solvable spin system  on co-evolving networks}

\author{Christoly Biely\inst{1},Rudolf Hanel \inst{1}, Stefan Thurner \inst{1,2} }

\institute{
\inst{1} Complex Systems Research Group; HNO; Medical University of Vienna; 
W\"ahringer G\"urtel 18-20; A-1090; Austria \\
\inst{2} Santa Fe Institute; 1399 Hyde Park Road; Santa Fe; NM 87501; USA 
}

\abstract{
We consider social systems in which agents are not only characterized by 
their states but also have the freedom to choose their interaction partners 
to maximize their utility. We map such systems onto an 
Ising model in which spins are dynamically coupled by links in a dynamical network. 
In this model there are two dynamical quantities
which arrange towards a minimum energy state in the canonical framework:
the spins, $s_i$, and the adjacency matrix elements, $c_{ij}$.
The model is exactly solvable because microcanonical partition functions reduce to products
of binomial factors as a direct consequence of the $c_{ij}$ minimizing energy. 
We solve the system for finite sizes and for the two possible thermodynamic limits and discuss
the phase diagrams. 
} 

\maketitle

\section{Introduction}

Properties of many statistical systems  are not solely  characterized by the states of their constituents, but
also depend crucially on how these interact with each other, i.e. their network (linking) structure. 
The way networks function can often not be fully understood by their linking structure alone 
because function may depend heavily on the internal states of individual nodes. 
Especially social and economical interactions are of this kind. Not only actions (states) matter but 
the possibility of choice with whom to interact (linking) plays a crucial role in socio-economical dynamics \cite{biely}.
It is therefore tempting to study the co-evolution of network structure and internal states.
In the simplest case, this can be done in the framework of the Ising model,
which immediately reminds of spin-glass models, such as the SK-model 
\cite{sherrington} or random-bond models, see e.g. \cite{idzumi}.
Ising models where both, spins and interactions, are governed by dynamical rules 
have been studied assuming different timescales of evolution, where typically 
interaction topology 'slowly' adapts in a pre-determined way on 'fast' relaxing spins \cite{coolen}.
Recently, such systems have been analyzed with the replica approach in the grand-canonical 
ensemble assuming that the interaction topology also minimizes the energy of the system 
\cite{allahverdyan}; the coupling of both, spins and interactions to heat-baths at different 
temperatures can be treated in the respective formalism as well \cite{allahverdyan2}. 
Note that these doubly-dynamical models are in marked contrast to the Ising model on {\em fixed} 
network structures, see e.g. \cite{doro_ising}. 
Complementarily the formation of network structure driven by various 
Hamiltonians has been investigated in some detail \cite{network_hamiltons}.
We think that a full understanding of many processes taking place in networks can only be 
achieved in a combined approach.
In the following we show that a spin system in the canonical ensemble 
where both, linking structure (given by the adjacency matrix $c_{ij}$)  
and spins $s_i$, minimize the energy, can be exactly solved 
since partition functions reduce to products of binomials. 

The following model is classically phrased in terms of magnetization of Ising spins. 
However, the main idea is that it can be one-to-one related to economic terminology. 
Magnetization, $m$,  correspond to  market shares in a situation of a two-company world. 
Think for example that there exist two telephone providers, A and B. The monthly cost for each 
individual  depends on its local connectivity (telephone call network) and on the costs per call  
(intra-provider and out of networks calls) fixed by the  
provider.  Here the state of an individual, $s_i$,  being customer of company A would relate to spin up, customers 
of firm B relate to spin down. Connectivity, $c_{ij}$, is determined by who calls whom. 
It is assumed that fully rational agents minimize their costs. The amount of rationality is  
modeled below by temperature, $T$. The external field, $h$, in the following relates to external biases, such as
asymmetries of PR activity in the firms.  There are no conceptual problms to extend the methodology of 
the present work to e.g. the Potts model, reflecting a more realistic situation of  multiple companies on the
market. 

\section{The model}

We  study  the Hamiltonian
\begin{equation}
  H(c_{ij},s_{i})=-J\sum_{i>j}c_{ij}s_i{}s_j-h\sum_{i}s_i  \quad,
 \label{gen_hamilton}
\end{equation}
where sums are taken over all $N$ nodes of the system. The position of
the links in the adjacency matrix $c_{ij}\epsilon\{0,1\}$ is a dynamical variable. 
The system has thus two degrees of freedom both minimizing energy:
the orientation of the individual spins $s_i\epsilon \{-1,1\}$  as usual, and the 
linking of spins, $c_{ij}$. $c_{ij}=1(0)$ means nodes $i$ and $j$ are (un)connected. 
We consider undirected networks ($c_{ij}=c_{ji}$), the
case of directed networks is a trivial extension  as pointed out below. 
We denote the number of spins pointing upward by
$n_\uparrow=\sum_i{} \theta(s_i)$, the number of links by $L=\sum_{i<j}c_{ij}$, 
magnetization $m=\frac{1}{N}\sum_{i}s_i=\frac{2n_\uparrow-N}{N}$, 
connectivity $c=\frac{L}{N}$, and connectedness $\varphi=\frac{L}{N^2}$. 
In the grand-canonical ensemble this Hamiltonian was studied in \cite{allahverdyan}
by use of the replica method.
In this work, we limit our interest to the canonical framework.

We start our analysis with the microcanonical partition function for energy $E$
\begin{eqnarray}
\Omega(E,N,L,h)&=&\sum_{\{c_{ij}\}}\sum_{\{s_{i}\}}\delta({H}(c_{ij},s_{i})-E) \nonumber \\
&=&\sum_{n_\uparrow=0}^N\Omega(N,n_\uparrow)\sum_{\{c_{ij}\}}\delta({H}(c_{ij},n_\uparrow)-E) \nonumber \\
&=&\sum_{n_\uparrow=0}^N\Omega(N,n_\uparrow)\Omega(E,N,L,h,n_\uparrow)
\quad,
\label{mic}
\end{eqnarray}
where $\Omega(N,n_\uparrow)$ is the number of  configurations
for a given $n_\uparrow$.  $\Omega(E,N,L,h,n_\uparrow)$ denotes the
microcanonical partition function for a fixed $n_\uparrow$.

In Eq.(\ref{mic}) the calculation becomes greatly simplified when  
realizing that a fixed number of spins pointing upwards, $n_\uparrow$, alone 
is sufficient  to determine the spin-state of the system since
one deals with all the different topologies for a given value of
$n_\uparrow$. 
In other words, the crucial observation is that 
the exact spin-configuration $\{s_i\}$ loses its relevance 
because the topology of the network is not fixed.
In this case partition functions simply reduce to binomial factors,  
\begin{equation}
\Omega(N,n_\uparrow)={{N}\choose {n_\uparrow}}
\quad , \quad 
\sum_{n_\uparrow=0}^{N}   {{N}\choose{n_\uparrow}} =2^N
\quad ,
\end{equation}
and the remaining  task is to determine $\Omega(E,N,L,n_\uparrow)$. 
To find the number of microstates leading to energy $E$ for fixed $n_\uparrow$, the only relevant physical fact is
whether a link $\ell$ connects two spins of (un)equal orientation, thus contributing a unit  $-J$ ($J$) to total energy.
The possible energy states are  
$  E\epsilon\{-LJ-Nhm,-LJ+2J-Nhm,...,LJ-2J-Nhm,LJ-Nhm\} $
where the lowest energy 
$-LJ-Nhm$ 
is realized if {\em all} links connect spins of equal orientation.
In general, if $k$ links connect spins of equal orientation ($L-k$ links connect spins 
of different orientation),  
$E=LJ-2kJ-Nhm$.
It is easy to see that the number of possible 'positions' of linking
spins of equal orientation, $a_e$, and unequal orientation, $a_u$, is given by 
\begin{eqnarray}
&a_e(N,n_\uparrow)=\frac{1}{2}(n_\uparrow(n_\uparrow-1)+(N-n_\uparrow)(N-n_\uparrow-1)) \nonumber \\
&a_u(N,n_\uparrow)=n_\uparrow(N-n_\uparrow) 
\quad , 
\label{nrpossibilities}
\end{eqnarray}
for undirected networks.
Directed networks trivially follow from
${a}_e^{\rm dir}(N,n_\uparrow)=2a_e(N,n_\uparrow)$ and 
${a}_u^{\rm dir}(N,n_\uparrow)=2a_u(N,n_\uparrow)$, because 
while in the undirected case, $0<L<N(N-1)/2$, in the directed case we have, $0<L^{\rm dir}<N(N-1)$.
Each link positioned in $a_{e(u)} (N,n_\uparrow)$ contributes  $-J$ ($J$) to the
total energy $E$.
Given Eq. (\ref{nrpossibilities}),  the microcanonical partition function for given $n_\uparrow$ and
the total partition function read
\begin{equation}
\label{micro_link}
\Omega(E,N,L,h,n_\uparrow)={{a_e(N,n_\uparrow)}\choose{\frac{(LJ-E-Nhm)}{2J}}}
{{a_u(N,n_\uparrow)}\choose{\frac{(LJ+E+Nhm)}{2J}} }\quad, 
\end{equation}
\begin{equation}
\Omega(E,N,L,h)=\sum_{n_\uparrow=0}^N
{{N}\choose{n_\uparrow}}
{{a_e(N,n_\uparrow)}\choose{\frac{(LJ-E-Nhm)}{2J}}}
{{a_u(N,n_\uparrow)}\choose{\frac{(LJ+E+Nhm)}{2J}}}.
\end{equation}

\begin{center}
\begin{figure}[ht]
\begin{center}
\resizebox{0.40\textwidth}{!}{\includegraphics{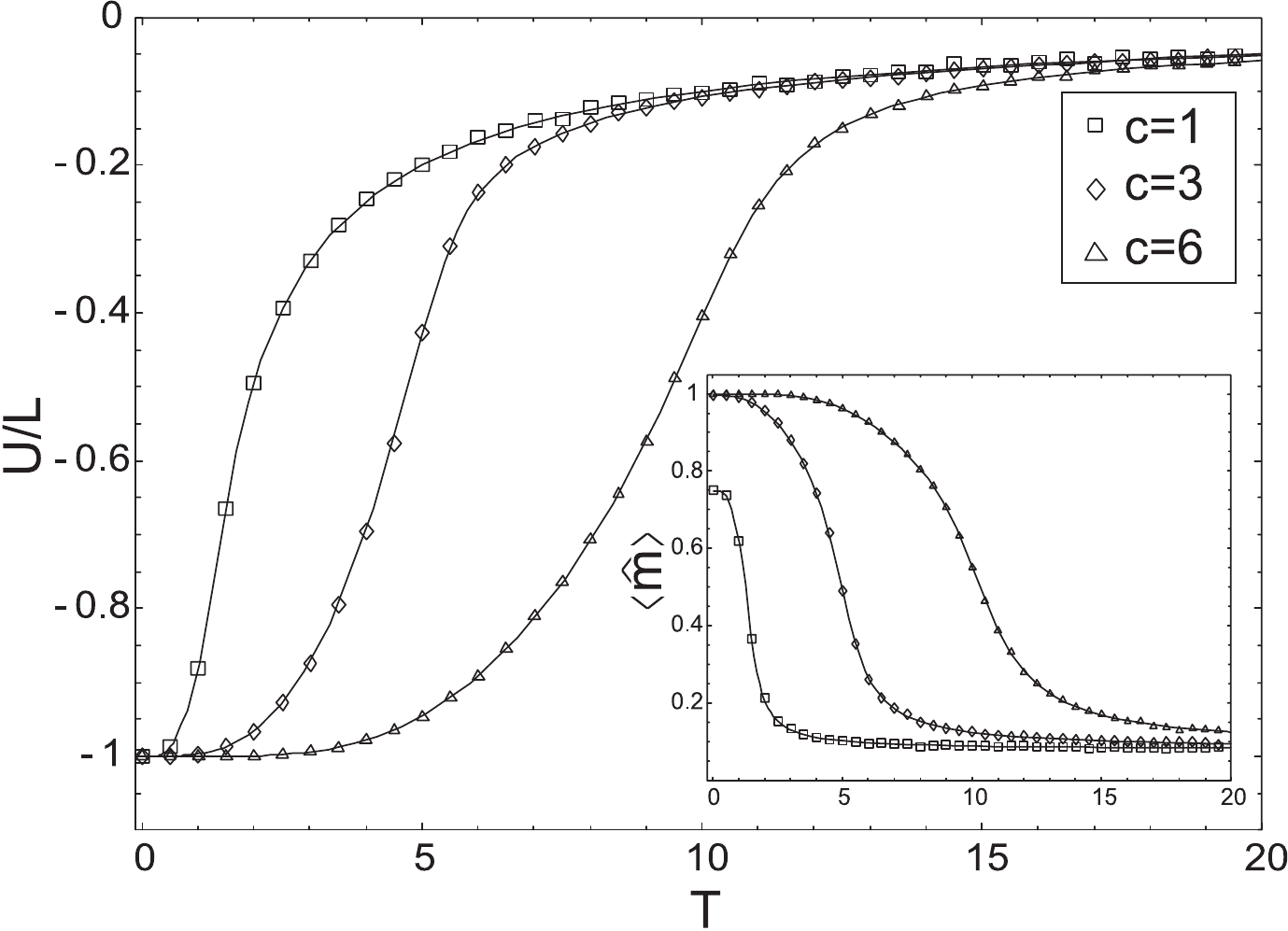}}
\end{center}
\caption{\label{pic1}
Internal energy  for $N=100$, and connectivities 
$c=1,3,6$. Solid lines correspond  to the exact finite size solution, Eq. (\ref{solution_exact}). 
Symbols are the results from a Monte Carlo simulation of the canonical ensemble pertaining to the
Hamiltonian of Eq. (\ref{gen_hamilton}). Inset: maximum of magnetization, $\bar m$ as a function of 
$T$ (lines exact, symbols MC). }
\end{figure}
\end{center}

We can now directly approach 
the problem of calculating the canonical partition
function $Z(T,N,L,n_\uparrow)$ of a system with fixed $n_\uparrow$ via
the Laplace transform, 
\begin{equation}
Z(\beta,N,L)=\sum_E\sum_{n_\uparrow}  {{N}\choose{n_\uparrow}} \Omega(E,N,L,h,n_\uparrow)e^{-\beta{}E}\quad. 
\label{Zex}
\end{equation}
Performing the energy summation the exact solution is
\begin{eqnarray}
Z(\beta,N,L,h,n_\uparrow)&=&e^{LJ\beta+Nhm\beta}\frac{\Gamma(1+a_e)}{\Gamma(1+L)\Gamma{(1+a_e-L)}}   \nonumber \\ 
 &\times&  {}_2\Phi_1(-a_u,-L,1+a_e-L,e^{-2J\beta}) \quad,  \qquad
\end{eqnarray}
with  ${}_2\Phi_1(-a,b,-c,x)=\sum_{k=0}^a{} \frac{(-a)_k(b)_kx^k}{(-c)_kk!}$ 
the hypergeometric function and  the Gamma function $\Gamma(x)$. 
The total canonical partition function finally is
\begin{equation}
\label{solution_exact}
Z(\beta,N,L,h)=\sum_{n_\uparrow=0}^N  {{N}\choose{n_\uparrow}} Z(\beta,N,L,h,n_\uparrow)
\quad , 
\end{equation}
and all thermodynamic quantities of interest are given exactly for finite sized systems, 
of (fixed) dimensions $L$ and $N$.
In Fig. \ref{pic1}  we show the internal energy $U$ and magnetization as a function of
temperature for different values of connectivity $c$ as calculated from
Eq. (\ref{solution_exact}). Perfect agreement with Monte Carlo simulations of
finite sized systems is found, where rewiring and spin-flipping have been implemented by 
the Metropolis algorithm. We note that for low connectivities the
obtained solutions are in very good agreement with the result of independent
spins, i.e. $U=c \tanh(\beta)$, as expected. 

\section{Thermodynamic limits}

\begin{figure}[t]
\begin{center}
\includegraphics[height=50mm] {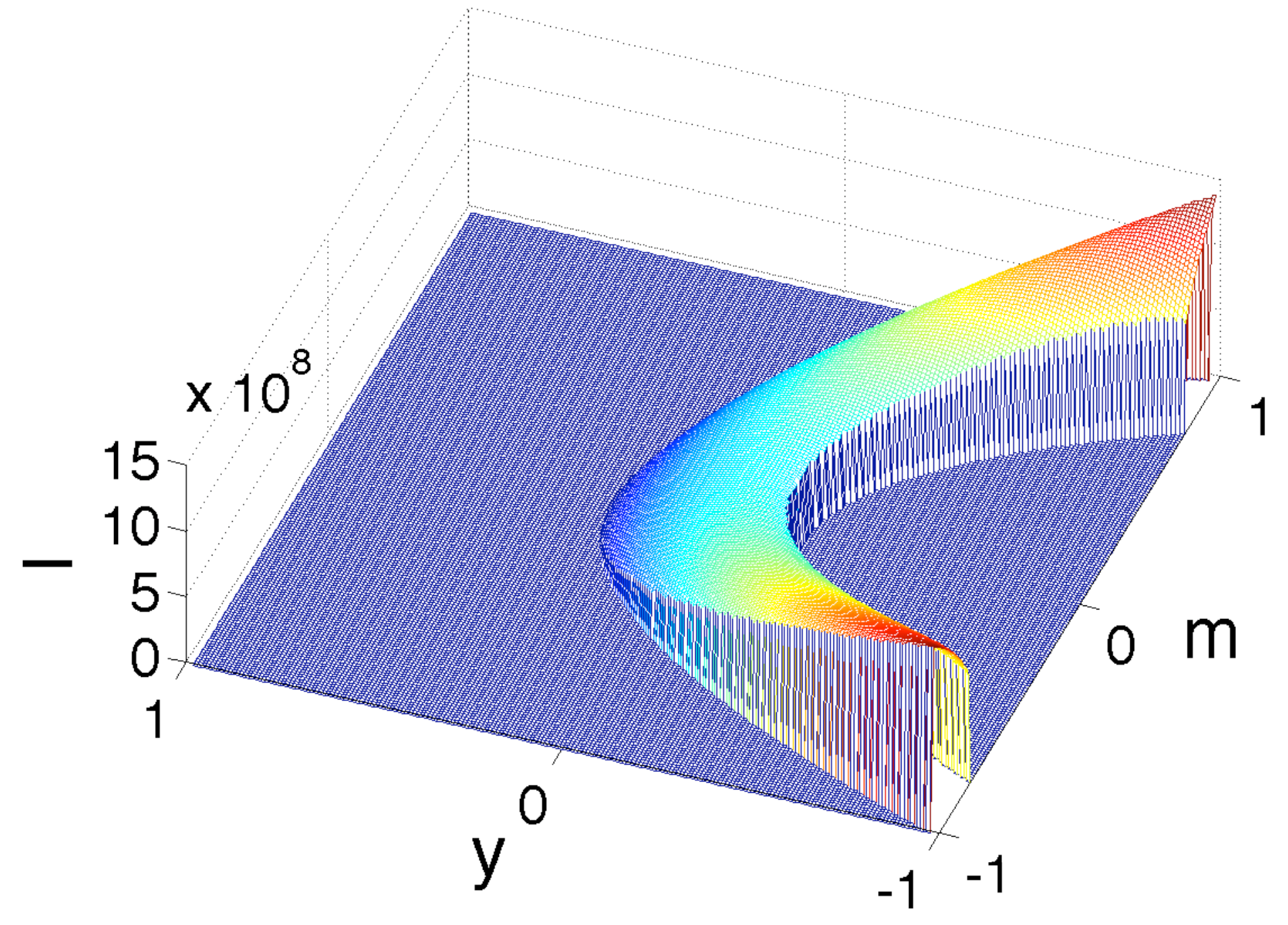}
\end{center}
\caption{Logarithm of Eq. (\ref{integrand}) in the $m-y$ plane for $\varphi=0.4$ and $c=20000$. 
The forbidden zones are clearly visible. The maximum is always reached in the allowed zone.}
\label{fig2} 
\end{figure}

Assuming large $N$, with Stirling's approximation in the form
${{a}\choose{b}}\sim \left[ (b/a)^{b/a}\right.$ $\left. (1-b/a)^{1-b/a} \right]^{-a}$,  
and the notation  $y=\frac{E}{LJ}$, Eq. (\ref{Zex}) reads
\begin{equation}
Z  =   2^{N-2} NL (2\varphi)^{-L} \int_{-1}^1dm \int_{-1}^1 dy  \left[ I(m,y,L,N)  \right] ^L \quad , 
\label{Z}
\end{equation}
with 
\begin{eqnarray}
&&I(m,y,L,N)=\exp(-\beta J y ) 
\left(  1-m^2 \right) ^{-\frac{1}{2c} }  \nonumber  \\
&&
\left(  \frac{1-m}{1+m}  \right)^{ \frac{m}{2c} }   
\left(  \frac{1-m^4}{1-y^2}   \right)^{ \frac{1}{2} } 
\left(  \frac{(1-y)(1-m^2)}{(1+y)(1+m^2)}   \right)^{ \frac{y}{2} } 
\nonumber \\ 
&&\times
\left( 1-2\varphi \frac{1-y}{1+m^2}  \right) ^{- \frac{1}{4 \varphi}(1+m^2-2\varphi(1-y) ) } 
\nonumber \\ 
&&\times 
\left(  1-2\varphi \frac{1+y}{1-m^2}  \right) ^{- \frac{1}{4 \varphi}(1-m^2-2\varphi(1+y) ) }  \quad , 
\label{integrand}
\end{eqnarray}
where $c$ and $\varphi$ are shorthands for $L/N$, and $L/N^2$, respectively; 
$h$ is set to zero for simplicity. In the thermodynamic limit $Z$ is reasonably approximated by the maximal 
configuration, i.e. the solution to  $d I / dy =0$, which is
\begin{equation}
y_{\rm max }= \frac{ -1-m^2t  +  \sqrt{(1+m^2t)^2 - 8 \varphi(m^2-(2\varphi-1)t )t  }  }   {4\varphi t } 
\label{maxi}
\end{equation}
with $t\equiv \tanh (\beta J)$. The other solution is outside the allowed parameter region of $y$. 
To ensure a real valued partition function the conditions, 
$1>2\varphi \frac{1-y}{1+m^2}$, and, $1>2\varphi \frac{1+y}{1-m^2}$, have to hold. Regions 
where they do not hold are forbidden zones in the 
$y-m$ plane, where the integrand 
of Eq. (\ref{Z}) is not defined, see Fig. \ref{fig2}. It can be 
shown that the maximum condition line, $y_{\rm max }(m)$, always stays in the allowed zone,
 $\forall$ $-1<m<1$, $0<t<1$, and $0<\varphi<1/2$. 

For the following discussion  let us compute the derivative of (the $\log$ of) $I$ in Eq. (\ref{integrand}), 
with respect to $m$  
\begin{eqnarray}
\label{dlogm}
&& 
\frac{d \log(I)}{dm}=\frac{m}{2\varphi } \log\frac{1-2\varphi\frac{1+y}{1-m^2} }{1-2\varphi\frac{1-y}{1+m^2} }
+\frac{1}{2c} \log\frac{1-m}{1+m} 
\nonumber \\
 &+&
 \left[\frac12
\log  \frac{2\varphi(1-y^2)  - (1-y)(1-m^2)}{2\varphi(1-y^2)  -  (1+y)(1+m^2)}  -  \beta J \right] y'  , 
\end{eqnarray}
with 
\begin{equation}
y'= \frac{m}{2\varphi }\left(   \frac{1+m^2 -4 \varphi }{\sqrt{(1+m^2 t)^2 -8\varphi(m^2-(2\varphi-1)t)t } } -1\right)
\quad.
\label{dummy}
\end{equation}

It is now natural to consider two distinct thermodynamical limits, $N\to\infty$, one, by keeping 
connectivity $c=L/N$, the other by keeping the connectedness $\varphi=L/N^2$, fixed.

\subsection{c=const. limit}

We fix $c$ and take $N \to \infty$. 
Consequently , $\varphi$ vanishes as $1/N$, and  
the maximum condition from Eq. (\ref{maxi}) reduces to 
\begin{equation}
 \lim_{\varphi\to 0 }  y_{\rm max }= - \frac{t + m^2}{1 + m^2t }   \quad .
\label{maxic}
\end{equation}
The limiting cases for infinite and zero temperature can be worked out immediately.

\subsubsection{The low temperature case, $\beta\gg1$}
\begin{figure}[t]
 \begin{center}
 \includegraphics[height=45mm] {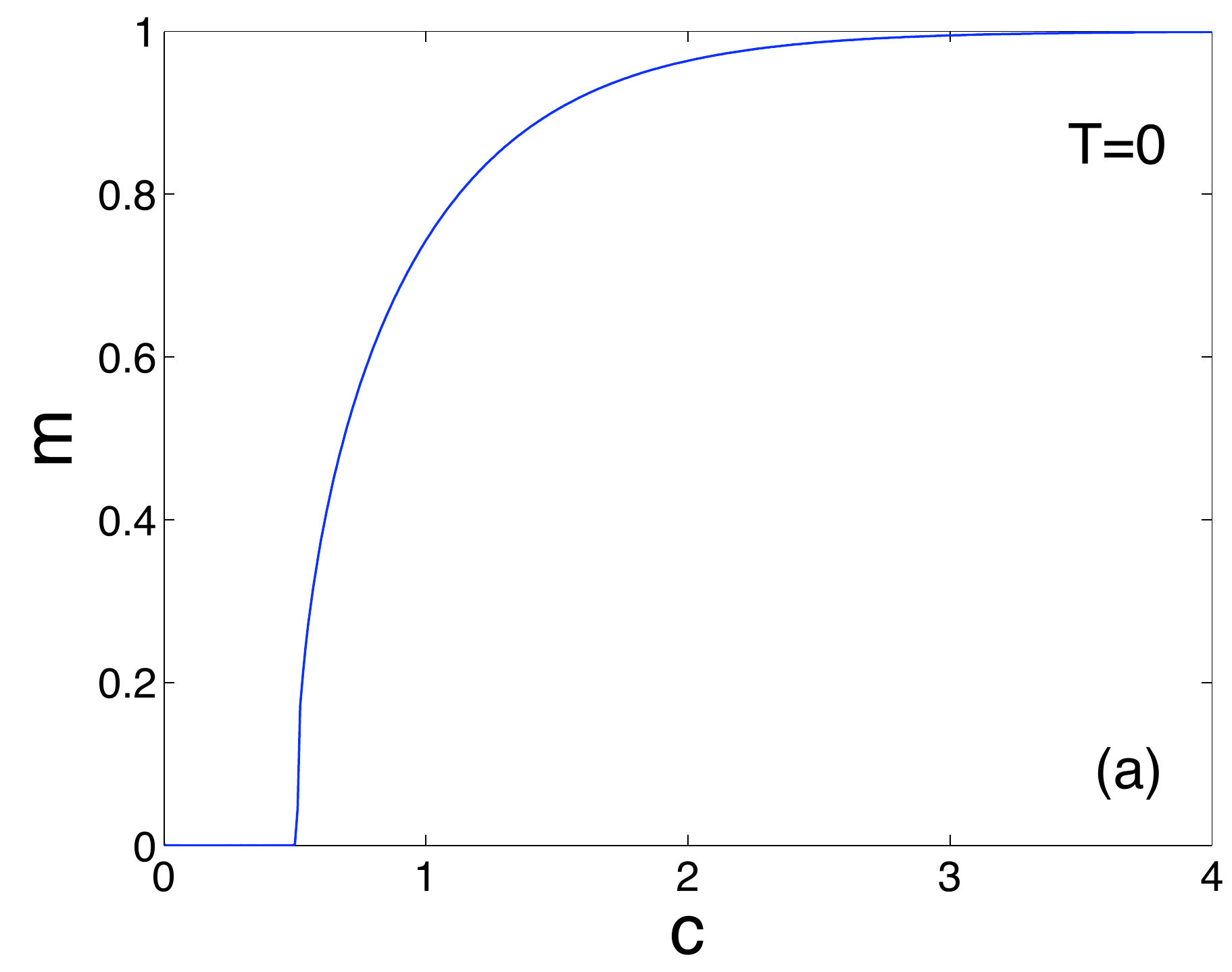}
 \includegraphics[height=45mm] {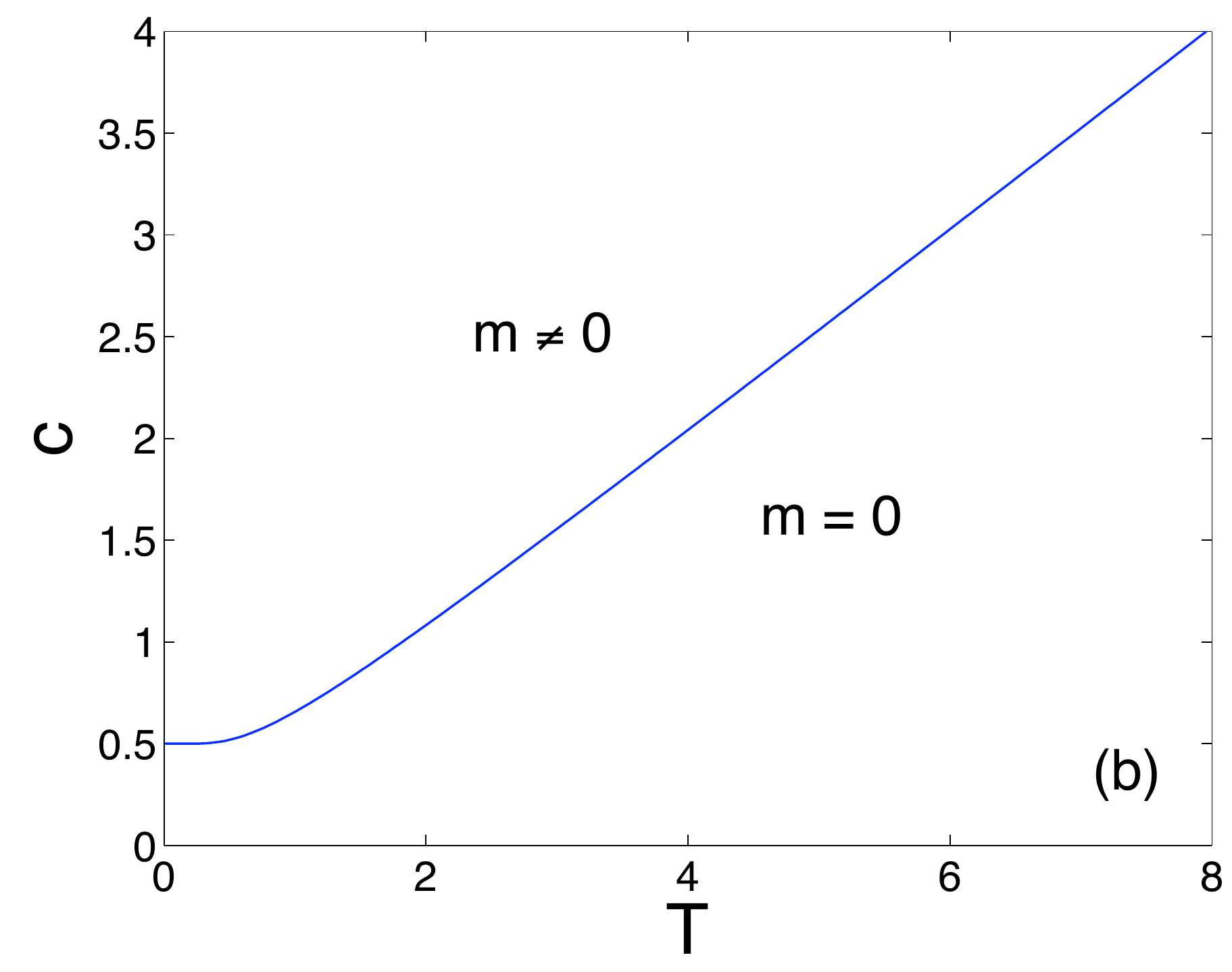}
 \end{center}
 \caption{
Fixed-$c$  thermodynamic limit. 
(a) Magnetization $m$ as a function of connectivity $c$ at zero temperature for $J=1$. 
Below $c=1/2$ there is no possibility for magnetization in the system. 
At $c\sim4$, practically full magnetization is reached. 
(b) Phase diagram in the $T-c$ plane for $J=1$. }
\label{fig3} 
\end{figure}

For $J\beta\gg1$, $t\to1$, and  the maximum condition further simplifies to, $y_{\rm max }=-1$, 
and $ y_{\rm max }' = \frac{2m}{1+m^2t}\left( \frac{t(t+m^2)}{1+ m^2t}  -1 \right)$. 
Using this in Eq. (\ref{dlogm}), and setting $\frac{d \log(I)}{dm}=0$ yields
\begin{equation}
\frac{1-m}{1+m} = \exp \left( -\frac{4cm}{1+m^2} \right)
\quad .
\label{sol1}
\end{equation}
The self-consistent solution  is shown in Fig. \ref{fig3}(a): We find zero magnetization below a
critical connectivity $c<1/2$, as well as a region where $m\neq0$. 
One can show \cite{allah_private} that Eq. (\ref{sol1})
can be obtained from the results pertaining to the grand-canonical ensemble studied in
\cite{allahverdyan} in the limit of an infinitely large chemical potential.

\subsubsection{Infinite temperature, $\beta\ll1$}
For $J\beta\ll1$, the maximum condition here becomes, $y_{\rm max }=-m^2$, and $y_{\rm max }'=-2m$. 
Proceeding as ,
from  $d \log(I)/dm =0$, we get 
$\log \left( \frac{1-m}{1+m} \right) =- 4c m \beta J$, or $m=\tanh(2cm\beta J)$. 
This implies no magnetization  for small $\beta$,  or $T\to\infty$, $\forall c$. 
The phase transition line, separating the phases $m=0$ and $m\neq0$, is found 
by noting in Eq. (\ref{integrand}) that for $c=$const. 
\begin{eqnarray}
&\lim_{N\to\infty } I(m,y,L,N)=e^{(1-\beta J y )} 
\left(  1-m^2 \right) ^{-\frac{1}{2c} }
\nonumber \\
&\times
\left(  \frac{1-m}{1+m}  \right)^{ \frac{m}{2c} }   
\left(  \frac{1-m^4}{1-y^2}   \right)^{ \frac{1}{2} } 
\left(  \frac{(1-y)(1-m^2)}{(1+y)(1+m^2)}   \right)^{ \frac{y}{2} } 
\quad .
\label{Ic}
\end{eqnarray}
Differentiating Eq. (\ref{Ic}) w.r.t. $m$, and setting it to zero, i.e. 
\begin{eqnarray}
0&=&\frac{1}{2c} \log\frac{1-m}{1+m} -2m\frac{y+m^2}{1-m^4} + \frac{m(t^2-1)}{(1+m^2t)^2} 
 \nonumber \\
&\times &
\left[ 
\log\frac{1-y}{1+y} +\log \frac{1-m^2}{1+m^2} -2 \beta J
\right]
\quad , 
\label{diff}
\end{eqnarray}
reduces for  $m\to 0$ to the critical line given by
\begin{equation}
\frac{1}{2c}= \tanh  \left( \frac{J}{T_{\rm c }}  \right)
\quad .
\end{equation}
Here we used 
$\lim _{m\to 0 } y_{\rm max } =\lim_{m\to 0}-\frac{m^2 +t}{1+m^2 t} = -t$. The phase diagram is 
shown in Fig. \ref{fig3}(b).  

Let us finish the $c=$const. case with a statement on the critical exponent $\beta$. 
By inserting  $y_{max}$, given in Eq. (\ref{maxic}), into Eq. (\ref{diff}), we get 
\begin{equation}
  t=\frac{\log \frac{1+m}{1-m}  }{ 4mc - m^2 \log \frac{1+m}{1-m} } \quad .
\label{te}
\end{equation}
Define $\tau= \frac{T_c-T}{T_c} > 0$, where $T=1/\beta$. Using the Ansatz $m=\lambda \tau^{\beta }$, where $\lambda$ is a free 
parameter,  in Eq. (\ref{te}), for $\tau \ll1$ one gets the expression
\begin{equation}
  \frac{12c \left( 1-\frac{1}{4c^2}  \right)}{1+\frac{1}{2c} } = \lambda^2 \tau^{2\beta-1 } \quad. 
\end{equation}
Since the left hand side does not depend on $\tau$, the only  possible choice  
for the critical exponent is 
$\beta=1/2$.  This is  exactly the mean field value.

\subsection{$\varphi=$const. limit}

Fixed $\varphi$ means diverging $c$,  for $N\to \infty$.

\subsubsection{The low temperature case, $\beta\gg1$}
For $J\beta\gg1$, the second term on the right hand side of Eq. (\ref{dlogm}) vanishes. 
A lengthy but trivial calculation shows that Eq. (\ref{dlogm}), without that term,  
is larger than zero for $0<m<1$, $\forall$ $0<\varphi<1/2$, and  $\forall$  $0<t<1$, where we used the fact that 
$y+m^2\leq0$. By symmetry, Eq. (\ref{dlogm}) (without the term $\sim1/c$) is negative for $-1<m<0$. 
This means there is no phase transition in the thermodynamic limit for $\varphi={\rm const. }$ 
and the system is always in a state of maximum magnetization, $m=\pm1$.

\subsubsection{The high temperature case, $\beta\ll1$}
For $J\beta\ll1$, we have $t=\tanh(J\beta)\sim J\beta$, and 
$y_{\rm max }=-m^2$, and $y_{\rm max }'=-2m$, as above. Setting 
$\frac{d \log(I)}{dm}=0$ in  Eq. (\ref{dlogm}), we get
\begin{equation}
m=\tanh(2m\beta J \varphi N)
\label{mag}
\quad, 
\end{equation}
which means $m=1$ in the thermodynamic limit. 
Note, that for the case where the coupling scales as $J=J_0/N$, we get a phase transition, 
for finite $J_0$, which is well known for the complete graph, i.e. $\varphi = 1/2$. 
In the context here, $\varphi$ shifts the phase transition.

More generally,  assume that the coupling scales with system size as $J=J_0 N^a$, with $a\le0$, i.e. coupling decreases with size. 
Assume further that $c$ scales as $c=c_0N^b$, with $b\ge0$, then $\varphi = c_0 N^{b-1}$. Using
Eq. (\ref{mag}) this means 
\begin{equation}
m=\tanh(2m\beta J_0 c_0 N^{a+b})
\label{mag2}
\quad .
\end{equation}
We see that  for $a=-b$, $\kappa=2\beta J_0 c_0$ can be seen as the critical parameter. For $a>-b$ we get $m=\pm1$, whereas 
for $a<-b$ there is no magnetization, $m=0$.

\section{Conclusion}

The crucial observation of this paper is that the summation over all topologies in the Ising model on dynamical networks
is equivalent to re-writing the partition function as a sum over all magnetizations. 
The model -- which can be seen as a toy model for a variety of socio-economical situations --  thus drastically reduces 
in complexity and becomes exactly solvable, both for finite size and the two possible thermodynamic limits. 


We thank A. Allahverdyan and an anonymous referee for various most useful and clarifying comments.
Supported by Austrian Science Fund FWF Projects P17621 and P19132.

\end{document}